# The News Delivery Channel Recommendation Based on Granular Neural Network

Lin Wu [1], Rui Li [2], Jiaxuan Liu [3,]*, Wong-Hing Lam [4]

[1] School of Computer and Cyberspace Security, Communication University of China; wulin@cuc.edu.cn
[2] School of Computer and Cyberspace Security, Communication University of China; lilrui@cuc.edu.cn
[3] School of Computer and Cyberspace Security, Communication University of China; schuan1999@163.com
[4] Department of Electrical and Electronics Engineering, University of Hong Kong; whlam@eee.hku.hk
* Correspondence: schuan1999@163.com; Tel.:(+86)18722188556

**Abstract:** With the continuous maturation and expansion of neural network technology, deep neural networks have been widely utilized as the fundamental building blocks of deep learning in a variety of applications, including speech recognition, machine translation, image processing, and the creation of recommendation systems. Therefore, many real-world complex problems can be solved by the deep learning techniques. As is known, traditional news recommendation systems mostly employ techniques based on collaborative filtering and deep learning, but the performance of these algorithms is constrained by the sparsity of the data and the scalability of the approaches. In this paper, we propose a recommendation model using granular neural network model to recommend news to appropriate channels by analyzing the properties of news. Specifically, a specified neural network serves as the foundation for the granular neural network that the model is considered to be build. Different information granularities are attributed to various types of news material, and different information granularities are released between networks in various ways. When processing data, granular output is created, which is compared to the interval values pre-set on various platforms and used to quantify the analysis's effectiveness. The analysis results could help the media to match the proper news in depth, maximize the public attention of the news and the utilization of media resources.

## 1. Introduction

With the rapid development of mobile Internet, people's lifestyles have undergone diverse changes, and many traditional industries have been impacted, resulting in new development patterns. News information are indispensable in people's daily life, where the news communication industry plays an important role in recording and spreading information by reporting current or recent news. As the speed of news updating accelerates and media forms become more diverse, the distribution of news channels has gradually become an issue. In recent years, the dissemination of news information has gradually shifted from the former mode of centralized release and dissemination centered on news media to the trend of terminal and platform development. Consequently, various internet-based news platforms have replaced traditional newspapers and magazines, becoming the primary means for the public to access news. Different news publishing platforms have distinct user demographics and preferences. These differences also lead to disparities in the attention received by similar news stories across different platforms. By comparing the attention received by various news articles on different platforms, it is possible to identify news publishing patterns that are better suited to a specific platform, effectively increasing the attention to the news platform and related news. The recommendation of news delivery channels is similar to the recommendation of users, but the difference is that there are more options for the delivery channels.

The current mainstream technical solutions are based on collaborative filtering and deep learning algorithms [1]. Collaborative filtering algorithms are currently mainly applied in the field of e-commerce recommendations [2]. They discover users' interests and preferences by mining their historical behavior data, and then divide users based on these interests and preferences, recommending similar news or items to users [3][4]. Collaborative filtering mainly has two types: user-based and item-based [5]. User-based algorithms recommend items that users with similar interests have liked but the target user has not purchased, while item-based algorithms recommend items similar to those that the target user likes [6]. Collaborative filtering algorithms make recommendations from the user's perspective, and the recommendation process is fully automated. The recommendations users receive are implicitly obtained from purchase patterns or browsing behaviors, without the need for users to actively seek suitable recommendations, such as filling out surveys, and without using personal information or product descriptions. The main steps in using collaborative filtering recommendation algorithms are to establish a user rating table, find similar users, and recommend items [7]. Collaborative filtering recommendation algorithms primarily face issues of data sparsity, cold start and robustness. When the rating matrix is sparse, the recommendation accuracy usually drops significantly. There is also the issue of recommendation efficiency in the big data environment. What's more, it can't generate suggestions for new items that have not received user ratings yet.

Deep learning is a new research field of machine learning research that studies how to automatically extract multi-level feature representations from data [8]. The basic idea is to extract features from data by combining low-level features to form denser high-level semantic abstractions, thus solving the problem of manually designing features in traditional machine learning [9]. For image and speech data containing a large amount of unlabeled data, deep learning models can learn more effective features from a large amount of unlabeled training data for recognition [10]. Deep learning could efficiently capture nonlinear and non-trivial user-item relationships, and is able to encode more complex abstractions into higher-level data representations, which has changed the architecture of recommendation algorithms, bringing more opportunities to improve the performance of recommendation programs [11]. However, the problem of deep learning is excessively redundant parameters. In order to achieve better network accuracy, a large number of parameters need to be set for the deep learning model and efficient optimization algorithm should also be considered [34]. Another drawback of deep learning-based recommendation algorithms is the difficulty of directly providing interpretable recommendation schemes, as the model output obtained through deep learning algorithm training is the weights between neurons in deep neural networks, making it difficult to directly provide reasonable explanations for the recommended results.

By analyzing the aforementioned methods, it can be found that the results of the collaborative filtering algorithm and the deep learning algorithm have a common flaw, both of them use a clear score to represent the user's preference, that is, the output of the algorithm is too single, corresponding to a specific value. However, user preferences are subjective and variable, clear scores cannot effectively measure the uncertainty of user preferences, which may affect the accuracy and efficiency of recommendation. In addition, in order to solve the problem of data sparsity, the algorithm in this paper mainly uses the combination of BP neural network and L1 regularization method and the combination of BP neural network and L1/2 regularization method to solve the problem. This method uses gradient descent method to update the weights and train the network, and finally gets a relatively stable result [12].

The news delivery channel recommendation algorithm based on granular neural network proposed in this paper can effectively solve the problem, at the same time improve the efficiency and influence of news delivery. We upgrade the traditional neural network, introduce the concept of granularity into it, then use the granularity neural network to screen the news delivery channels. Since the output of the granular neural network is a numerical interval, which is wider and closer to the real situation than the previously determined value range. The main innovations of this paper are as follows:

- Instead of analyzing specific user behaviors, different media resource user profiles are constructed, and the distribution proportion of news is determined based on core users. The difference between the news distribution channel recommendation based on granular neural networks and traditional recommendation system algorithms lies in the target audience. The traditional recommendation system algorithm mainly aims at the historical behavior and preference information of users, while the news delivery channel recommendation based on granular neural network is for different media platforms;
- Granular neural networks generate a granularity interval, replacing specific numerical values with interval values, and comparing them with the pre-defined interval values of different media platforms, thus allowing for more reasonable and effective news distribution channel recommendations [13][14];
- In different media platform, which could use interval number in order to gain more comprehensive recommendations based on the results of innovation, we can complete another innovation point，that is a one-to-many or many-to-many mapping relationship, which is due to the different news delivery channel user portraits can be learned through the mature and efficient algorithm, and there are often multiple users who choose the same news delivery channel. In this way, the one-to-many or many-to-many mapping between the binary relation group of news and users can be realized, and the efficiency and correctness of deep learning, that is, granular neural networks, can be greatly improved.

The rest of this paper is organized as follows: Section 2 overviews the current recommendation algorithms. Section 3 introduces the main technical methods of the granular neural network and the recommendation of news delivery channels based on the granular neural network combined with the objective function. In Section 4, experiments are conducted based on the constructed news dataset and algorithms. Finally, Section 5 summarizes the work of this paper. The overall structure diagram of this paper is shown in Figure 1.

*News channel recommendation based on granular neural network*

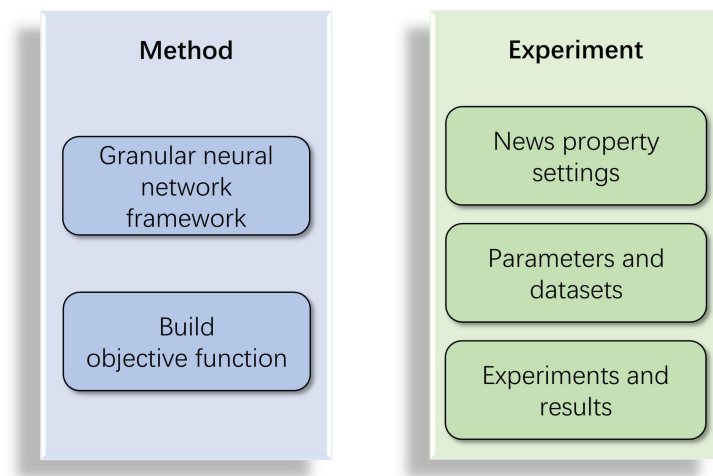

**Figure 1.** The overall structure diagram.

## 2. Related Works

The current mainstream recommendation algorithms mainly include collaborative filtering algorithm and deep learning recommendation algorithm.

*2.1. Collaborative filtering*

Collaborative filtering algorithm has always been one of the most popular and applicable recommendation algorithms [15]. with its ideas closely aligned with real-life situations. The algorithm's performance outperforms most other general algorithms in most cases, and the recommendation results are highly likely to meet users' personalized needs, making it a key research subject. Meng et al. proposed an improved dynamic collaborative filtering algorithm called Hybrid Dynamic Collaborative Filtering (HDCF), which is based on topic models [16]. They added a time decay function to the topic model, which assigns different weights to items based on the time users view them. the approximate variational posterior value of the model is given, and its variational inference model is also available. Wu et al. proposed that user preferences are subjective and variable, the specific scores cannot effectively measure the uncertainty of user preferences, and proposed an interval-valued triangular fuzzy scoring model [17]. Based on the user score statistics, this model replaces specific scores with interval value triangular fuzzy numbers, which can more reasonably measure user preferences. The optimized algorithm calculates user similarity using interval value triangular fuzzy numbers and considers the fuzziness of scores in the prediction stage. However, this algorithm faces some difficulties in finding the optimal parameter k, requiring further research to design a model that automatically determines parameters.

*2.2. Deep learning*

Deep learning-based recommendation systems typically take various user and item-related data as input, use the related models to learn the latent representations of users and items, and generate item recommendations for users based on these latent representations [18][19]. According to the types of data used in the recommendation systems and combined with the classification of traditional recommendation systems, current research can be mainly divided into five directions:

- Content-based recommendation systems using deep learning [20]. By utilizing explicit or implicit user feedback, user profiles, item content, and other forms of user-generated information, deep learning methods are employed to learn the latent vectors of users and items, and recommend similar items to users [21];

- Applying deep learning techniques to collaborative filtering. By using explicit or implicit user feedback, deep learning is employed to learn the latent vectors of users or items, and then predict user ratings and preferences based on these latent vectors;
- Incorporating deep learning techniques into hybrid recommendation systems. By combining explicit and implicit user feedback, user profiles, and item content data, recommendations are made using different content generated by different users. At the model level, content-based recommendation algorithms and collaborative filtering algorithms are combined [22];
- Applying deep learning in social network-based recommendation systems [23]. Based on various data, such as explicit or implicit user feedback and users' social relationships, deep learning models primarily model the influence of social relationships among users, better discovering users' preferences for items;
- Utilizing deep learning techniques in context-aware recommendation systems. These systems use various types of data, such as explicit or implicit user feedback, and employ deep learning models to establish users' environmental preferences.

He et al. proposed a recommendation algorithm integrating multilayer perceptron (MLP), which is a model used to solve the XOR problem. By adding hidden layers to the neural network and having fully connected layers for both hidden and output layers, MLP can be used to add nonlinear transformations to existing recommendation methods and interpret them as neural extensions [24]. In most cases, recommendations are considered as bidirectional interactions between user preferences and item features. For example, matrix factorization decomposes the rating matrix into low-dimensional user and item latent factors, making it convenient to construct a dual neural network to simulate the bidirectional interactions between users and items.

## 3. Methods

This paper analyzes the multi-channel delivery of different kinds of news based on granular neural network. Through the comparison of a large amount of data, this paper finds and summarizes some hidden laws and characteristics of different news delivery platforms, and utilizes these characteristics to make effective suggestions for each news delivery platform to help the platform optimize the category and content of news delivery. We introduce the traditional granular neural network system, then systematically describe the network structure and training process of granular neural network, and the objective function of combining coverage and specificity proposed on this basis, finally analyze the insufficiency of the objective function determined in this way from different angles. Then we put forward the concept of balance degree, which improve the performance of granular neural network. At this time, the objective function is determined by the combined effect of coverage, specificity and balance degree. The particle size in the granulation process can be determined more appropriately by using this objective function. The algorithm needs to quantify news topics according to feature classification, and then granulate the values according to the granularity generated by the objective function to obtain the final result interval [25]. Compare this interval with the range corresponding to the multi-channel media until the corresponding media is found, then the media is the most suitable distribution channel for the news [26].

*3.1. Granular Neural Network Framework*

The structure of a neural network is similar to that of an equation set with approximate solutions. The data set used for training is the parameters at the corresponding positions of the equation set, and the training is equivalent to the process of finding approximate solutions [27], as shown in Figure 2. After training, we could obtain the network model of the neural network, and the approximate solution obtained is the parameters of the network model, and this set of parameters corresponds to the overall model of the neural network [28]. The granular neural network generated by the fusion of granular computing and neural network, through the combination of genetic algorithm, particle swarm optimization and other heuristic optimization algorithms, successfully completed the experiment of reflecting realistic parameters from granular neural network [29]. In the traditional multiple linear regression task, the multiple linear model can be obtained by training the data set, but it is impossible to judge which attribute of the data set is more important through the parameters of the model. The granular neural network that integrates granular computing optimizes the granularity allocation through heuristic algorithms, and can distinguish the importance of different attributes [30].

Formally, the network model of the granular neural network only has one more step of data granulation than the common BP (Back Propagation) network model. In addition, the more difference lies in the granulation method and the internal calculation of the granular neural network. In order to achieve the goal of selecting important attributes of the data set, the granulation method adopts the method of intervalizing the attributes of the data set, and the calculation within the granular neural network adopts the method of interval computing [31]. After the granulation method is clarified, as shown in Figure 3, the network structure of the granular neural network is given, in which the input layer is the input original data, that is, the data set obtained after the quantization of news attributes, and the granulation layer is the data after granulation. , the results of the output layer are used to participate in the calculation of the objective function of the heuristic optimization algorithm, thereby optimizing the granularity allocation that guides data granulation. On the other hand, from the perspective of inequalities or micro-elements, it is obvious that operations such as addition, subtraction, multiplication and division of interval calculations can be obtained.

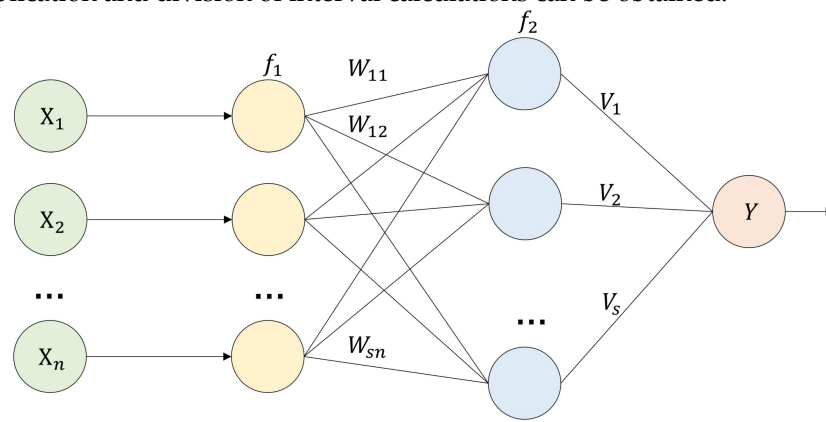

**Figure 2.** The traditional neural network with hidden layers.

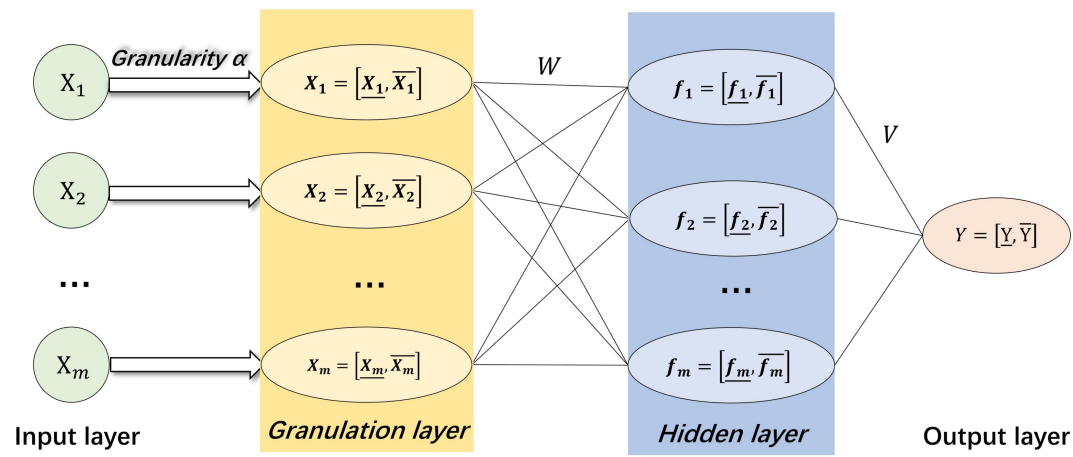

**Figure 3.** Granular neural network with granular input features and structure.

The text continues here (Figure 2 and Table 2). The granular neural network is formed on the basis of a given neural network, but the granular neural network and the neural network are not in a competitive relationship, it is established on the basis of the existing neural network [32]. Granular connections are formed in the form of intervals, the distribution and quantification of information granularity is the key content of granular neural networks [33]. The granularity of information is distributed among the connections of the neural network in various ways, in order to maximize some specific performance indicators [35][36]. The allocation protocol is high-dimensional and can be optimized by a variety of optimization algorithms, such as particle swarm optimization, single-parent genetic algorithm, etc. [37]. Granular neural networks focus on the granular output of the network, which must be evaluated against the numerical goals of the data, taking into account two criteria: the coverage of the numerical data and the specificity of the information granularity. Any number input into a granular neural network would produce a granular output, i.e., a specific interval range [38]. The training process of granular neural network could be divided into the following steps. First, the traditional BP neural network model is trained with the data set, then the granularity is allocated to each attribute in the data set, the data is granulated, and finally the relatively optimal granularity allocation strategy is found through the heuristic optimization algorithm [39], while the granularity of each attribute allocation reflects the importance of the attribute. When the granularity of an attribute allocation is smaller, the importance of the attribute is higher [40]. Its flow chart is shown in Figure 4.

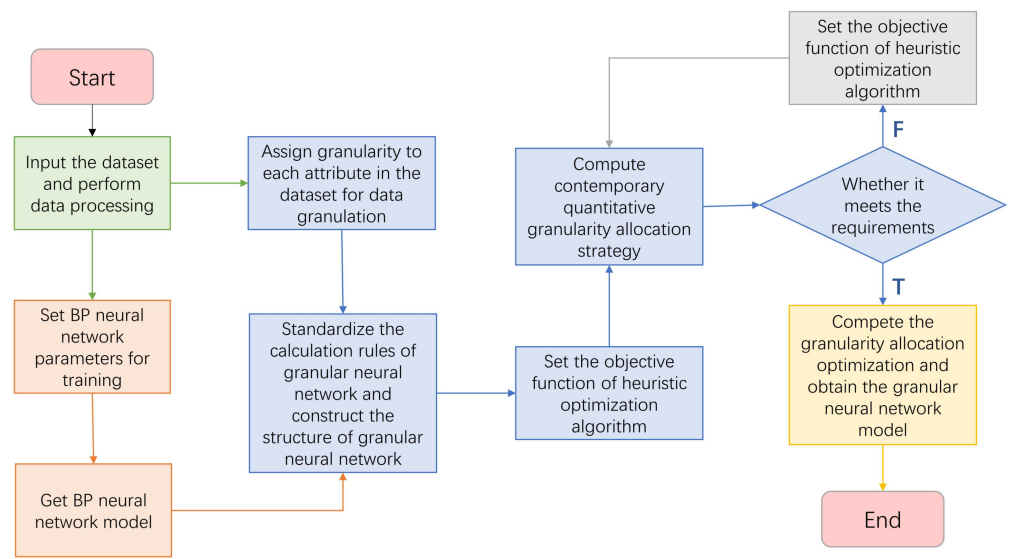

**Figure 4.** The training flow chart of granular neural network.

First, set up a number of news release platforms. At the same time, classify and input THUCNews data sets containing several types according to news topics, set BP neural network parameters, and obtain the BP neural network model. Second, the granularity of each news attribute of different categories in the data set is allocated, and the data is granulated. Combined with the obtained BP neural network model, the granular neural network model structure is constructed. Finally, set the objective function of the heuristic optimization algorithm and calculate the granularity allocation strategy. If the granularity allocation meets the requirements, complete the allocation. If the granularity allocation does not meet the requirements, continue to iterate the algorithm until the granular neural network model is obtained. After that, the interval vector of different news attributes output by the granular neural network is compared with the preset interval of each news platform, and the most suitable news platform for this news attribute is obtained.

*3.2. Build the objective function*

In addition, based on the granular neural network obtained by neural network training, and considering different optimization methods, this paper proposes an objective function that combines coverage, specificity and balance. Furthermore, we apply the optimized granular neural network to the recommendation of news delivery channels based on granular neural network. The objective function is used because the output of the model is an interval, but usually the objective of the experimental data is based on specific values, so an appropriate performance indicator must be defined, and the optimization of the performance indicator can be achieved through appropriate information granularity allocation, here the performance indicator is the objective function [41].

The determination of the objective function is a difficult problem. In our existing research, we propose the concepts of coverage and specificity. Among them, coverage refers to the statistical index that each output in the data set falls into the interval range obtained by the corresponding input through the model; specificity refers to the measurement index of the average width of the interval obtained by the model for all inputs. In order to give the formulas of these two indicators, we assume that the data set D has a total of n pieces of data, and each piece of data has m attributes and one output. In the process of granularity allocation, we set the granularity for the system as $\alpha$, that is, the initial granularity of each attribute of the data is $\alpha$. For coverage, we get the following formula, where $n0$ refers to the number of data set outputs fall into the corresponding interval, and n refers to the total data volume of the data set:

$$C = {n_0}/{n} \qquad (1)$$

Similarly, for specificity, the following formula can also be given to express the average width of the model output interval under unit granularity:

$$S = \frac{\sum_{i=1}^{n}(Z_{iup} - Z_{ilow})}{n \cdot m \cdot \alpha} \qquad (2)$$

For the entire model, when the coverage rate is higher and the specificity is lower, the results of the model not only fit the original data set output, but also fit more accurately, and the distribution of granularity is more reasonable. However, these two concepts are complementary and contradictory to each other. We expect that the lower the specificity, the more reasonable the granularity distribution is, but the premise is that the interval of the model output could cover the original output of the data, otherwise it violates the principle of regression. At the same time, we cannot increase the coverage rate unconditionally, because when the coverage rate is always one, it is difficult to determine whether the specificity at this time is still instructive.

The degree of balance that plays an important role in the objective function is a concept proposed for the unbalanced distribution results. We expect that by introducing this indicator in the objective function, the results of granularity allocation tend to be balanced. Its realization principle comes from the mean value inequality $(a_1 + a_2 + \ldots + a_n)/n \geq (a_1 \cdot a_2 \cdot \ldots \cdot a_n)^{\frac{1}{n}}$. To make the inequality hold, it is necessary to ensure that the value of $a_1 + a_2 + \ldots + a_n$ is a constant value, which is the same as the purpose of this model.

Assuming that the data set D has m attributes, the granularity of the i-th attribute is $\alpha_i$, where i equals to 1, 2, …, m. According to the model requirements, $\alpha_i \geq 0$, so $\alpha_1 \cdot \alpha_2 \cdot \ldots \cdot \alpha_m \geq 0$. And because $\alpha_1 + \alpha_2 + \ldots + \alpha_m$ is a constant value, so $0 \leq \alpha_1 \cdot \alpha_2 \cdot \ldots \cdot \alpha_m \leq [(\alpha_1 + \alpha_2 + \ldots + \alpha_m)/m]^m$, the equal sign on the right is if and only if $\alpha_1 = \alpha_2 = \ldots = \alpha_m$.

Back to the granular neural network model, in order to make the granularity distribution tend to be balanced, the most ideal state we expect is $\alpha_1 = \alpha_2 = \ldots = \alpha_m$, which just makes $\alpha_1 \cdot \alpha_2 \cdot \ldots \cdot \alpha_m$ reach the maximum value, and we least expect seeing that the granularity of an attribute is 0, that is, it is the least desirable to see $\alpha_1 \cdot \alpha_2 \cdot \ldots \cdot \alpha_m$ achieve the minimum value. Therefore, due to the high degree of conceptual agreement, $\alpha_1 \cdot \alpha_2 \cdot \ldots \cdot \alpha_m$ is suitable for representing balance degree.

However, in practical applications, it is found that when a certain $\alpha i$ is small, the value of $\alpha_1 \cdot \alpha_2 \cdot \ldots \cdot \alpha_m$ would also become smaller, which makes it difficult to play a certain role when it constitutes an objective function with coverage and specificity. Therefore, it is necessary to deal with the current balance to a certain extent. The most appropriate method is to use the $(x_0 - x_{min})/(x_{max} - x_{min})$ formula to normalize it. From this, the balance can be summarized as the following formula:

$$B = \frac{\alpha_1 \cdot \alpha_2 \cdot \ldots \cdot \alpha_m - 0}{\left(\frac{\alpha_1 + \alpha_2 + \ldots + \alpha_m}{m}\right)^m - 0} \qquad (3)$$

The following formula can be obtained by sorting, where $\alpha$ refers to the granularity given by the system, i.e., the initial granularity of each attribute:

$$B = \frac{\alpha_1 \cdot \alpha_2 \cdot \ldots \cdot \alpha_m}{\alpha^m} \qquad (4)$$

In the process of integrating the balance degree into the objective function, the value of the balance degree can be flexibly adjusted in the final objective function through the vertex form of the one-dimensional quadratic function with the help of the vertex characteristic of the one-dimensional quadratic function. Now the objective function $Q_1$ is given as follows:

$$Q_1 = \begin{cases} C \cdot \frac{1}{e^S} \cdot (-(B-b_0)^2 + (1-b_0)^2), 0 \leq b_0 \leq \frac{1}{2} \\ C \cdot \frac{1}{e^S} \cdot (-(B-b_0)^2 + b_0^2), \frac{1}{2} \leq b_0 \leq 1 \end{cases} \quad (5)$$

In the above formula, $b_0$ is the value of the balance degree we expect. Let the unary quadratic function $f(B) = -(B-b_0)^2 + X$, from the properties of the unary quadratic function, $f(B)$ will take the maximum value when $B = b_0$, and the vertex value X can guarantee that when the independent variable is [0,1], $f(B)$ would not take a negative value. The proof is as follows: when $0 \leq b_0 \leq \frac{1}{2}$, the symmetry axis of $f(B)$ is between 0 and 0.5, as long as $f(1) = 0$ is guaranteed, $f(B)$ is always greater than or equal to 0 in $\left[0, \frac{1}{2}\right]$, at this time $X = (1-b_0)^2$. Similarly, when $\frac{1}{2} \leq b_0 \leq 1$, it is necessary to ensure that $f(0) = 0$, at this time $X = b_0^2$.

Combined with the above objective function, the improved granular neural network algorithm is put into the application of news delivery channel recommendation, which could effectively improve the efficiency of the algorithm.

## 4. Experiments

### 4.1. News property settings

After several steps of improvement, the theory is applied to the experiment of multichannel news delivery. Before officially starting the experiment, we give 11 possible attributes according to the topic of the news report, as shown in the following table:

**Table 1.** News attributes corresponding to news topics.

| News attributes | Attribute content |
|---|---|
| The current politics news | News related to the Party and government guidelines and policies issued, the convening of meetings and other current political news |
| The economic news | News related to production, circulation, distribution, consumption and other social and economic life |
| The culture and travel news | News related to culture and tourism industry |
| The sports news | News related to competitive sports and related events, sports training |
| The social news | News about social events and issues related to People's Daily life and their interests |
| The rule of law news | News concerning the establishment, enforcement and supervision of the legal system |
| The science and technology news | News related to the field of science and technology and used to disseminate scientific knowledge to the general public |
| The education news | Related to education policies and regulations, educational research results, recruitment |
|  | information, education issues of all ages and other aspects of the news |
| The military news | News about war, military building, soldiers' lives and armed conflict |
| The international news | News about hot issues in other countries |

Then perform interval value judgment and attribute evaluation on the corresponding media channels. Taking Tencent News, Sina Weibo and Toutiao as examples, Tencent News, as one of Tencent's industries, has 'social networking' as its biggest feature. It builds a more complete ecosystem centered on social networking, which facilitates the collection of user data and brings together a large number of data media resources. Compared with other social media, Sina Weibo is more inclined to be a social media, and it is closer to the real society in terms of social information dissemination. Sina Weibo currently has the largest social content sharing and dissemination ecosystem, and plays a pivotal role in social dissemination. Different from the above two media, Toutiao is not a news client in the traditional sense. Its core operation is a set of algorithms built from code. The algorithm model records every behavior of the user on Toutiao, calculates the user's preferences, and pushes the content that the user is most likely to be interested in. Toutiao is building its own digital content ecosystem by capturing massive amounts of Internet information, content aggregation for Toutiao authors, and current strategic investment in the short video field.

*4.2. Parameters and datasets*

The python version used in this experiment is 3.6, the corresponding Anaconda version is 4.3.0, and the corresponding pytorch is 1.7.0.

We need to set the interval value of each news delivery channel. The specific parameter interval value is shown in the following table:

**Table 2.** Interval value setting of news distribution channels.

| Channel of news delivery | The interval numerical |
|---|---|
| People's Daily Online, Xinhua News Agency | [9,10] |
| CAIJING.com.cn | [7,9] |
| China Cultural Tourism Network | [6,7] |
| Tencent News, Toutiao | [5,10] |
| www.legaldaily.com.cn | [3,4] |
| zol.com.cn | [2,4] |
| www.JYB.cn | [3,12] |
| www.81.cn | [1,2] |
| www.huanqiu.com, www.haiwainet.cn | [1,6] |
| weibo.com | [3,10] |

Compare the number of attribute value intervals of the news obtained from the experiment with the value interval of the channel, then choose the channel with the most suitable and highest matching degree to deliver news.

After giving the above 11 news attributes and platform value intervals, use the THUCTC dataset (a subset of the THUCNews news text classification data set provided by the Tsinghua NLP group) to apply the data set to the experiment. THUCNews is generated by filtering and filtering the historical data of Sina News RSS subscription channels, including 740,000 news documents, all in UTF-8 plain text format. 70% data for training, 30% data for testing.

*4.3. Experiment steps and results*

The granularity is determined for the dataset by the degree of balance in the objective function Q, which is calculated by the following formula:

$$B = \frac{\alpha_1 \cdot \alpha_2 \cdot \ldots \cdot \alpha_m - 0}{\left(\frac{\alpha_1 + \alpha_2 + \ldots + \alpha_m}{m}\right)^m - 0} \tag{6}$$

The data set D has m attributes, and the granularity of the i-th attribute is $\alpha_i$. After iterative optimization of granularity, the following experimental results are obtained. Similar to the THUCNews dataset mentioned in this paper, we use THUCTC, a text classification tool, which is a Chinese text classification toolkit launched by the Natural Language Processing Laboratory of Tsinghua University. It can automatically realize the training, evaluation and classification functions of user-defined text classification corpora. The difference with THUCNews is that we do not obtain the classified news text that can be directly used in the model, but need to use the classification function of THUCTC to preprocess the massive news that is manually crawled [42]. Because the THUCNews already contains the relevant data of sina, so we change the different websites, including kwangmyong, People's Daily has a certain degree of authority of the news media's official website, to ensure the accuracy and authenticity of the news, avoid false news causing interference with the result of the experiment, to reduce the noise from the source. It can be seen from the results that THUCNews still performs better and can minimize the influence of input on the final result of the algorithm 43. Therefore, we only show the experimental results of THUCNews here.

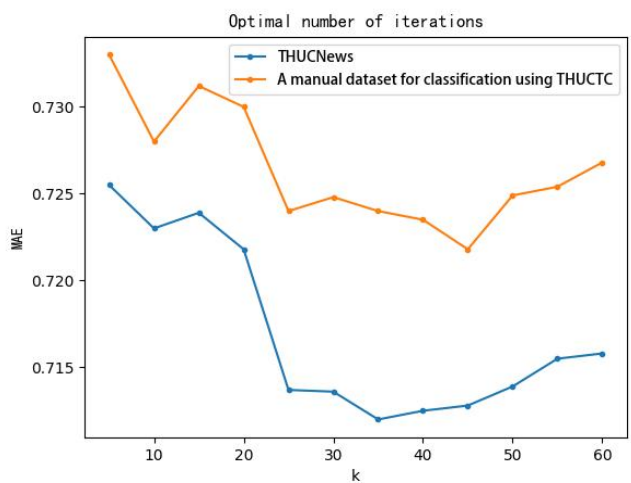

**Figure 5.** Optimal granularity.

As can be seen from the above figure, the final optimal particle size should be *α*=0.5. After determining the optimal granularity, it is necessary to granulate the data set, and obtain the value interval of the news attribute after rounding. The specific interval numerical results are as follows:

**Table 3.** Numeric range of news attributes.

| News attribute | current affairs | economy | Cultural tourism | rule of law | technology | military |
|---|---|---|---|---|---|---|
| Numerical interval | [9,10] | [7,8] | [6,7] | [3,4] | [2,3] | [1,2] |

| News attribute | education | internationality | sports | society | entertainment |
|---|---|---|---|---|---|
| Numerical interval | [4,10] | [2,6] | [3,9] | [5,10] | [3,9] |

Finally, match the numerical range with the delivery channel to obtain the final platform selection result.

In the experiment, the performance of the algorithm is quantified by the commonly used mean absolute error (MAE), which is used to measure how close the prediction is to the true result, calculated by the following formula:

$$MAE = \frac{1}{N}\sum_{i=1}^{N}|P_i - R_i| \qquad (7)$$

where $P_i$ is the predicted value, $R_i$ is the true result, and N is the length of the list. The lower the MAE is, the higher the accuracy would be. At the same time, the algorithm results are compared with the collaborative filtering algorithm based on cosine similarity, i.e., using cosine similarity to calculate user similarity. In order to reasonably estimate the user rating mode, by setting the parameters and adjusting the value range, according to the results in the figure below, it can be concluded that THUCNews reaches the best performance when k=34, which is 0.7121.

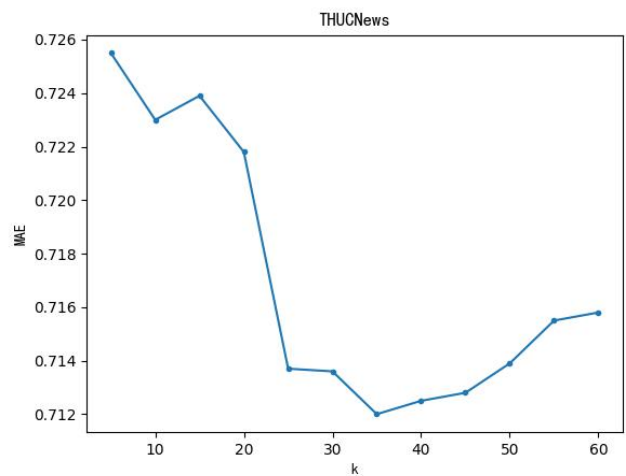

**Figure 6.** The optimal number of iterations k.

On the basis of the improved granular neural network, that is, on the basis of the granular neural network combined with the proposed objective function, the following experiments are carried out, which are compared with the collaborative filtering algorithm based on cosine similarity. Based on this, we set the parameters for the collaborative filtering algorithm based on cosine similarity, and use the best k to conduct a comparative experiment, and the results in the following figure can be obtained:

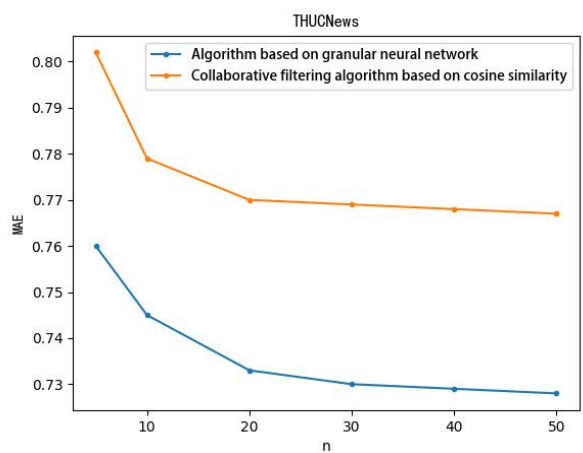

**Figure 7.** Comparison of the two algorithms.

where n is the number of times parameter in the two algorithms. From this, it can be concluded that the algorithm based on granular neural network could generate a numerical interval, which has higher accuracy and wider application range than the collaborative filtering algorithm based on cosine similarity that generates specific numerical values.

On the basis of using the THUCTC data set and determining the granularity $\alpha$ = 0.5, Figure 8 shows the relationship between the mean absolute error (MAE) and the number of hidden neurons of the data set. The main trend is when the number of hidden neurons increases, the MAE value decreases. However, the curve itself has been fluctuating. This is because the performance of the neural network is affected by the initialization of the weights and biases (with the same learning rate and the same training data.)

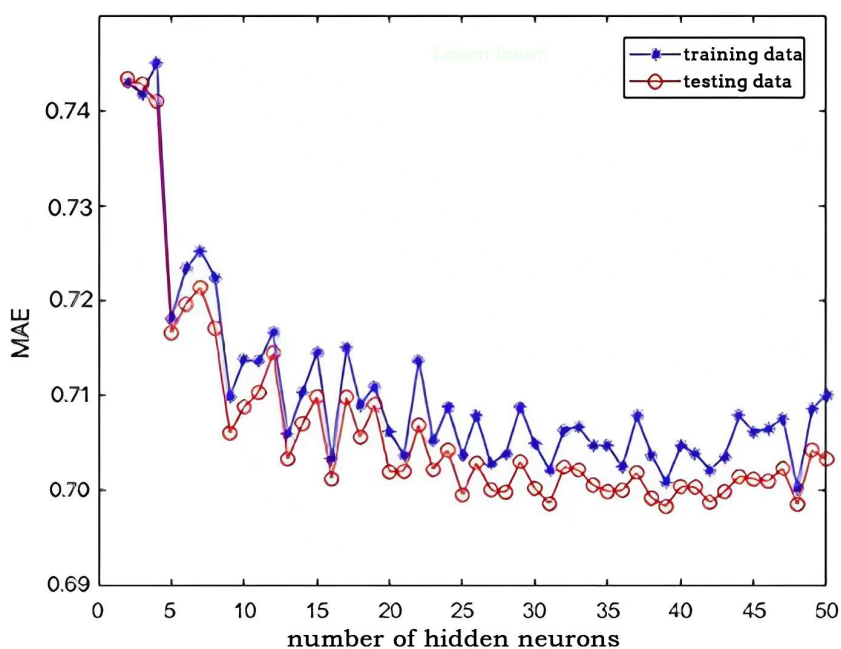

**Figure 8.** MAE versus the number of hidden neurons in the data set.

Through the above experiments, this paper draws the following conclusions: 1. News delivery should pay attention to media positioning and responsibilities. In particular, news with strong professionalism and high threshold for collecting and writing is more in line with traditional media organizations and vertical media. The experimental results found that current affairs news is suitable for posting on the People's Daily Online and Xinhua News Agency media platforms, economic news is suitable for posting on CAIJING.com.cn, cultural tourism news is suitable for publishing on China Cultural Tourism Network, legal news is suitable for posting on www.legaldaily.com.cn, science and technology news is suitable for posting on zol.com.cn, education news is suitable for posting on www.JYB.cn, military news is suitable for posting on www.81.cn, and international news is suitable for posting on www.huanqiu.com and www.haiwainet.cn. 2. Social and entertainment news and online media platforms are more compatible. The emerging network media has formed relatively significant user habits. News with lower professionalism and reading thresholds and closely related to public interests is more suitable for posting on online media platforms. The experiment results show that sports news and social news are suitable for posting on Toutiao, Tencent News and other platforms; entertainment news is posted on weibo.com.

## 5. Conclusions

To utilize a granular neural network-based news distribution channel recommendation algorithm, it is necessary to analyze the news attributes based on the granular neural network model and complete the task of granularity allocation. Different granularities are assigned to news items with different attributes, quantifying news attributes. In this step, an objective function combining coverage, specificity, and balance is used, with any numerical input into the granular neural network producing granular output.

Subsequently, news platforms are analyzed and studied, using classical algorithms to construct user profiles for different media platforms and assigning different granularity interval values to these platforms. The output of the granular neural network is compared with these interval values to investigate hidden patterns and characteristics within various news distribution platforms.

The research on granular neural network-based news distribution channel recommendation algorithms can enhance the deep matching between media and news, optimizing news distribution channels. This algorithm requires pre-setting media platforms, and the future research direction lies in exploring how to recommend news distribution channels without specifying platforms in advance. It is worth mentioning that our proposed method is able to scale in large scale networks, and our future work will mainly focus on extending it in federated learning scenarios [43][44].

## 6. Patents

**Author Contributions:** Conceptualization, Lin Wu; methodology, Rui Li.; software, Jiaxuan Liu.; validation, Rui Li; formal analysis, Jiaxuan Liu; data curation, Wong-Hing Lam; writing—original draft preparation, Rui Li; writing—review and editing, Jiaxuan Liu. All authors have read and agreed to the published version of the manuscript.

**Funding:** This work is supported by the science and technology project of State Grid Corporation of China (The key technologies and applications of brand influence evalution and monitoring fits into the layout of "One body and four wings",grant no.1400-202257240A-1-1-ZN).

**Data Availability Statement:** Not applicable.

**Conflicts of Interest:** The authors declare no conflict of interest.